\documentclass[fleqn,twoside]{article}
\usepackage{espcrc2}

\usepackage{graphicx}
\usepackage{epsfig}
\usepackage[figuresright]{rotating}

\newcommand{\AmS}{{\protect\the\textfont2
  A\kern-.1667em\lower.5ex\hbox{M}\kern-.125emS}}

\title{Ultra-High Energy Cosmic Rays: The Annihilation of Super-Heavy Relics}

\author{\underline{P. Blasi}
	\address[arcetri]{Osservatorio Astrofisico di Arcetri\\ 
        Largo Enrico Fermi, 5 - 50125 Firenze, ITALY}
        \thanks{blasi@arcetri.astro.it}
	R. Dick\address[canada]{Department of Physics and Engineering Physics,
        University of Saskatchewan,\\116 Science Place, Saskatoon, 
        SK S7N 5E2, Canada}
	and 
	E.W. Kolb\address[fermilab]{NASA/Fermilab Astrophysics Center,
        Fermi National Accelerator Laboratory,\\ Batavia, Illinois 60510-0500}
	\address[chicago]{Department of Astronomy and Astrophysics,
        Enrico Fermi Institute,\\ The University of Chicago, Chicago, Illinois 
        60637-1433}}
       
\begin{document}

\begin{abstract}
We investigate the possibility that ultra-high energy cosmic rays (UHECRs) 
originate from the annihilation of relic superheavy (SH) dark matter in the
Galactic halo.  In order to fit the data on UHECRs, a cross section of 
$\langle\sigma_Av\rangle\sim 10^{-26}\textrm{cm}^2 (M_X/10^{12}\,
\textrm{GeV})^{3/2}$ is required if the SH dark matter follows a 
Navarro--Frenk--White (NFW) density profile.  
This would require extremely large-$l$ contributions to the annihilation cross
section.  An interesting finding of our calculation is that the annihilation 
in sub-galactic clumps of dark matter dominates over the annihilations in the
smooth dark matter halo, thus implying much smaller values of the cross
section needed to explain the observed fluxes of UHECRs.
\vspace{1pc}
\end{abstract}

\maketitle

\section{Introduction}

The detection of particles with energy in excess of $10^{20}$ eV is one
of the unsolved mysteries of modern astrophysics and might represent one
of the many signals of the existence of physics well beyond the standard
model of particle physics. In \cite{BKV,KR} it was proposed the appealing 
possibility that this mystery might be related to another problem, the
origin of dark matter. In fact the dark matter might be made of SH relics
of the big bang \cite{ckr1,ckr2,ckr3,ckr4} with lifetimes possibly 
exceeding the age of the universe. 
The rare decays of these SH particles (with masses in excess of $10^{12}$ GeV) 
naturally generates UHECRs from the top, that is as end-product of their
decay. In order to saturate the dark matter content of the universe and
at the same time explain the fluxes of UHECRs, the lifetime of the SH relics
should be of order $10^{22}$ years, which is difficult to achieve for such
massive particles unless some discrete conserved symmetry is introduced that
prevents the natural decay of the relics on much shorter time scales.

In this paper we investigate an alternative possibility, consisting in the 
annihilation of SH relics rather than in their decay. 
The simple assumption that dark matter is a thermal relic limits the maximum
mass of the dark matter particle to less than a few hundred TeV.
However, it has been recently pointed out that dark
particles might have never experienced local chemical equilibrium during the
evolution of the universe, and that their mass may be in the range $10^{12}$ to
$10^{19}$ GeV, much larger than the mass of thermal WIMPs
\cite{ckr1,ckr2,ckr3,ckr4}. 

In this paper we calculate the expected flux of UHECRs from the smooth dark 
matter content of our Galaxy, and the flux due to clumps of dark matter in
the halo.

\begin{figure}[htb]
\includegraphics[width=8cm,height=4cm]{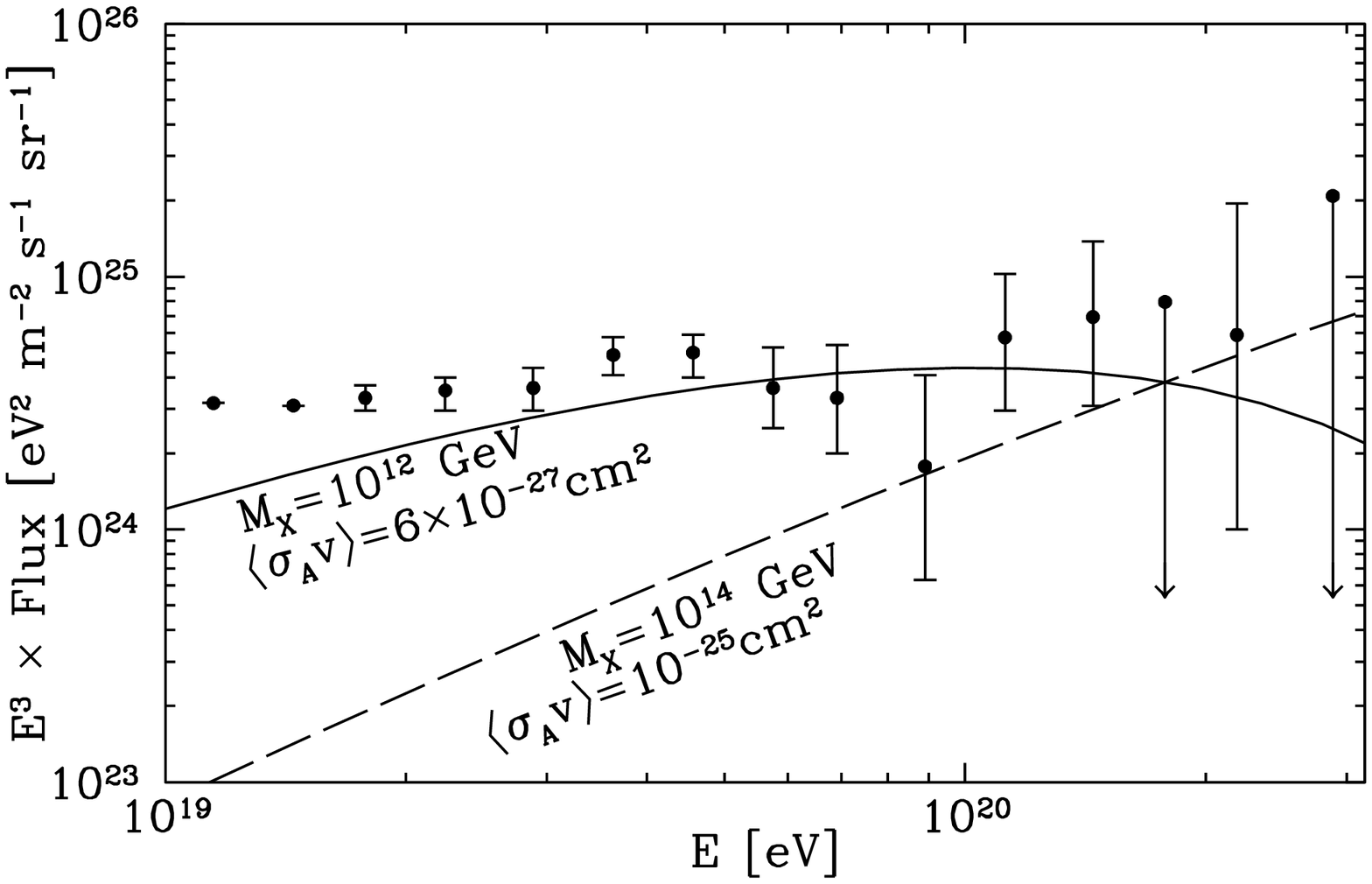}
\includegraphics[width=8cm,height=4cm]{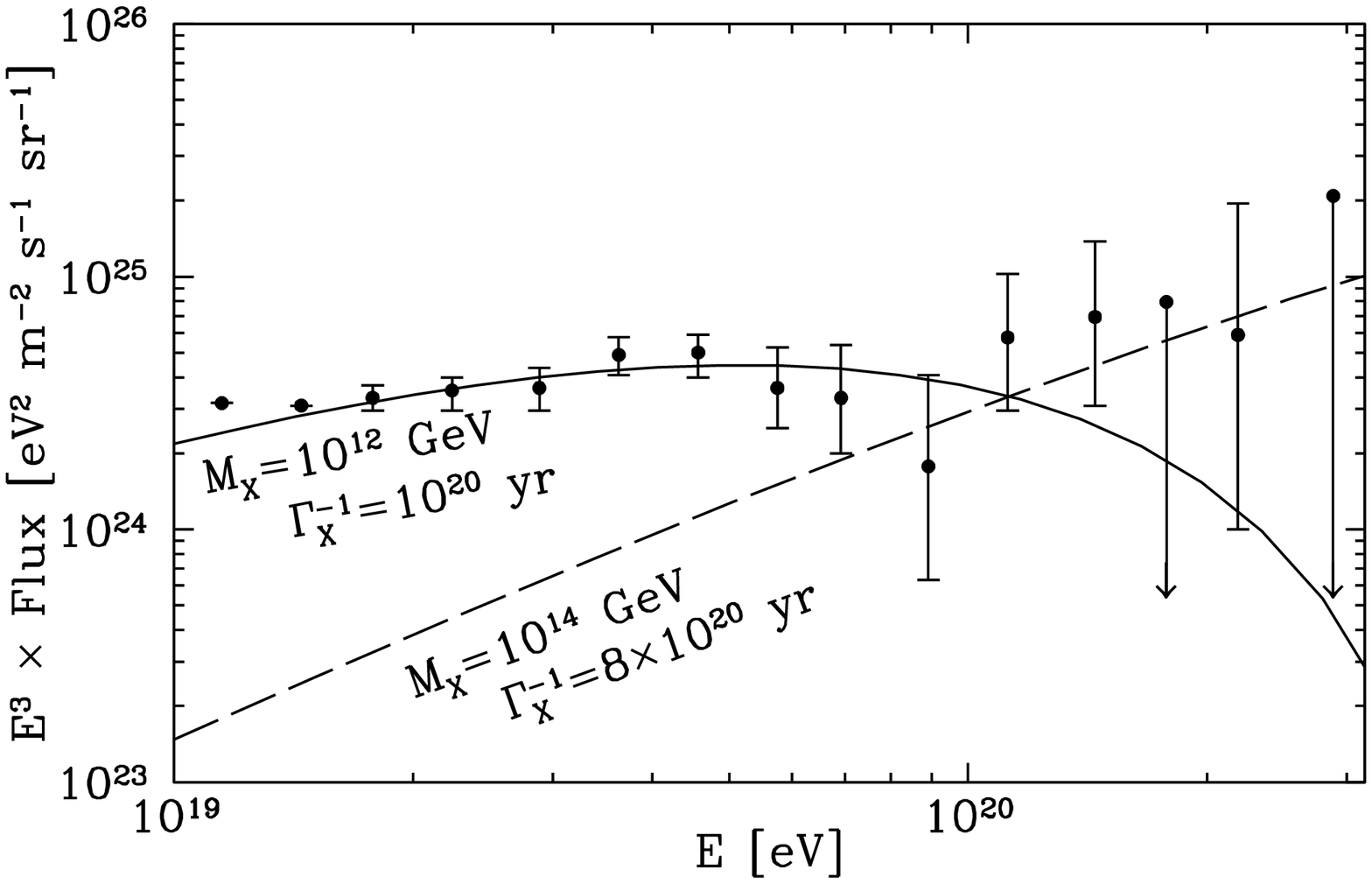}
\caption{UHECR spectra from SH particle annihilation (upper
panel) or decay (lower panel).  For both figures the solid (dashed) lines 
are for $M_X=10^{12}\,$GeV ($M_X=10^{14}\,$GeV). Cross sections and decay rates
are indicated in the plot.}
\label{spectrum}
\end{figure}

\section{SH Dark matter annihilation and UHECRs}

In calculating the UHE cosmic ray flux from a smooth
SH dark matter distribution in the galactic halo,
we assume a halo density spherically symmetric about
the galactic center, in the form of a NFW profile \cite{NFW}
$n_X(d)=\frac{N_0}{d(d+d_s)^2}$. We will use $d_s = 3d_\odot = 24\,$kpc in our
numerical estimates, where $d_\odot\simeq8\,$kpc is the distance of the solar
system from the galactic center.
The dimensionless parameter $N_0$ may be found by requiring that the total mass
of the Galaxy is $2\times 10^{12} M_{\odot}$.
For simplicity, we will assume that each annihilation produces two
quark initiated jets, each of energy $M_X$, while the decay of a SH particle  
produces two jets, each of energy $M_X/2$. The energy spectrum of 
observed UHE cosmic ray events from annihilation is determined by the 
energy distribution of the particles in each jet, which we take to equal
the fragmentation function in the modified leading 
logarithmic approximation (MLLA) \cite{MLLA}, which was also
employed in \cite{BKV,bbv}.

We also use the MLLA limiting spectrum in the results of Fig.\ \ref{spectrum}. 
Calculating the resulting UHE cosmic ray flux in the annihilation model, and
comparing it to the similar calculation in the decay scenario, we obtain the
results shown in Fig.\ \ref{spectrum} (the data points are from AGASA).
In order to provide enough events to explain the observed UHE cosmic rays,
$\langle\sigma_Av\rangle$ has to be in the range $10^{-25}\,\textrm{cm}^2$ to
$10^{-27}\,\textrm{cm}^2$, which is well in excess of the unitarity
bound to the $l$-wave reaction cross section \cite{SW,GK,hui}.
The unitarity bound essentially states that the annihilation cross section must
be smaller than $M_X^{-2}$.  However, as emphasized by Hui \cite{hui}, there
are several ways to evade the bound.  The annihilation cross section may be
larger if there are fundamental length scales in the problem larger than
$M_X^{-1}$.  

A related issue is the typical energy of the annihilation products.  In this
paper we assume that annihilation produces two jets, each with energy
approximately $M_X$.  It is easy to imagine that with the finite-size effects
discussed above, there is the possibility that annihilation will produce many
soft particles, rather than essentially two particles each of energy $M_X$.
An example that suits our needs is the annihilation of a monopole-antimonopole
pair, as discussed at length in \cite{astrop}.
We regard the requisite size of the annihilation cross section to be the most 
unattractive feature of the annihilation scenario.

This problem becomes less severe when the clumped component of the dark
matter distribution in the Galactic halo is taken into account. The 
spatial and mass distribution of the clumps is taken from numerical 
N-body simulations (see \cite{astrop} and \cite{Sheth}). The presence of 
cuspy density profiles in the clumps makes their contribution to the diffuse
flux of UHECRs dominant over the NFW smooth component. The ratio of the clumped
to smooth contributions for isothermal and NFW profiles for the clumps and
for different values of $M_X$ is plotted in Fig. 2.
\begin{figure}[htb]
\includegraphics[width=7cm,height=5cm]{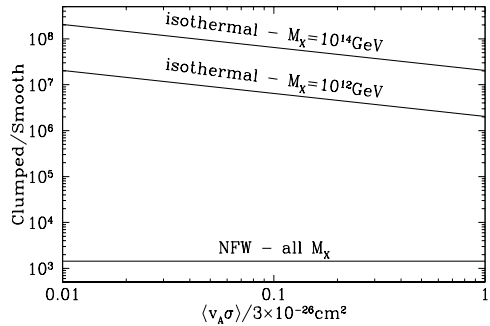}
\caption{The ratio of events from the subclump component to the smooth
component assuming either an isothermal or a NFW profile for the subclumps.}
\label{ratio}
\end{figure}
In the case of isothermal clumps, where the density profile of a single clump
scales like $1/r^2$, the main contribution to the annihilation comes from 
the inner regions of the clumps, so that it is crucial to define a minimum
radius $R_{min}$, determined by the annihilation efficiency. In \cite{astrop} 
it is explained in detail the effect of two different definitions for 
$R_{min}$.

Another peculiar prediction of the annihilation of SH relics in clumps of
dark matter in the halo consists of a peculiar pattern of small scale 
anisotropies. Several multiple events within degree scales would be observed
if annihilation in the clumps generates the UHECRs. A detailed simulation
was carried out in \cite{astrop}. The model also predicts a large scale 
anisotropy in the direction of the galactic center, which is an important 
test to be carried out by the future experiments, that will be able to see the
direction of the galactic center.

\section{Conclusions}

Our calculations can be summarized in the following points: 1) the 
annihilation scenario for the origin of UHECRs requires cross sections 
that may be larger that the unitarity bound. Although not a killing factor, 
for the reasons explained above, this is certainly an unappealing feature 
of the model; 2) the annihilations in clumps of galactic dark matter dominate
over the contribution of a smooth NFW dark matter profile, requiring
correspondingly smaller values for the annihilation cross section; 3) the
pattern of anisotropies is such that both a large scale (dipole) anisotropy
and a degree scale anisotropy is to be expected if UHECRs are generated in 
the annihilations of SH relics in galactic SH dark matter.

\end{document}